\documentclass{iopconfser}
\usepackage{amssymb,amsmath,amsthm,amsbsy,epsfig,color,graphicx,times,physics}
\usepackage[ansinew]{inputenc}
\usepackage[english]{babel}
\usepackage{color}
\usepackage{braket}
\usepackage{tabularx}
\usepackage{array}
\usepackage{hyperref}
\usepackage{subcaption}
\usepackage{slashed}
\usepackage{enumitem}
\usepackage{float}

\begin{document}

\title{Searching for mirror neutrons and dark matter with cold neutron interferometry}

\author{Antonio Capolupo$^{1}$, Gabriele Pisacane$^{1}$, Aniello Quaranta$^{2}$ and Francesco Romeo$^{1}$}

\affil{$^1$Dipartimento di Fisica ``E.R. Caianiello'' Universit\`{a} di Salerno, and INFN -- Gruppo Collegato di Salerno, Via Giovanni Paolo II, 132, 84084 Fisciano (SA), Italy}
\affil{$^2$School of Science and Technology, University of Camerino,
	Via Madonna delle Carceri, Camerino, 62032, Italy}

\email{capolupo@sa.infn.it, gpisacane@unisa.it, aniello.quaranta@unicam.it, fromeo@unisa.it}

\begin{abstract}
We report a novel neutron interferometry scheme aimed at probing the potential existence of mirror neutrons, which have been proposed as viable dark matter candidates. Our theoretical analysis demonstrates that if mirror neutrons exist, ordinary neutrons would acquire a measurable geometric phase as a result of their mixing with these mirror counterparts \cite{NMN1}. 
\end{abstract}

\section{Introduction}

Neutrino oscillations \cite{Neut1,Neut2,Neut3,Neut4,Neut5,Neut6,Neut7,Neut8,Neut9,Neut10}, anomalies in particle physics such as the muon $g-2$ \cite{Muon1,Muon2} and the neutron lifetime discrepancy \cite{Czarnecki,Mumm,Berezhiani2019}, as well as the observed presence of dark matter and dark energy \cite{Rubin1,Rubin2,Trimble1987,Corbelli2000,Clesse2018,Capolupo2021,Salucci2021,Capolupo2020, Capolupo2025} motivates the research for physics beyond the Standard Model. Notably, the theoretical framework developed to explain one phenomenon often helps to explain another. An example is the axion, which simultaneously can solve the strong CP problem \cite{Peccei1,Peccei2,Weinberg,Wilczek,Raffelt,CapolupoAx2020,Marsh} and serves as a candidate for dark matter. Several extensions of the Standard Model, including supersymmetry and mirror matter theory \cite{YangLee1956,Kobzarev1966,Blinnikov1982,Berezhiani2004,Hodges1993,Foot2014,Hao2022,Hostert2023,Berezhiani2006}, predict new particles that could both restore fundamental symmetry in nature and account for the missing matter in the Universe. Mirror matter can manifest itself through neutron-mirror neutron mixing \cite{Berezhiani2019,Hostert2023,Berezhiani2006}, a phenomenon we investigate here.
Neutron interferometry \cite{RauchBook,Werner1979,Wagh1990,Allman1997,Filipp2005,CapolupoAx2021} and atomic interferometry \cite{brax} have proven to be powerful techniques for probing fundamental particle interactions and their quantum properties. In particular, neutron interferometry has allowed the experimental verification of several theoretical predictions, including the existence of the Berry phase \cite{Berry,Berry1} and the non-cyclic geometric phase \cite{Mukunda}.\\
We show that a geometric phase \cite{Geo1,Geo2,Geo3,Geo4,Mukunda,Capolupo2018} arises naturally for the neutron as a consequence of its mixing with its mirror counterpart. We report on a cold neutron interferometry experiment designed to detect this effect. Furthermore, we further analyse how the proposed experimental setup can probe a substantial region of the parameter space, potentially imposing  constraints on the mirror matter model and providing direct evidence for this hidden gauge sector.

\section{Neutron-mirror neutron mixing}
In order to restore parity symmetry, the concept of a hidden gauge sector was first introduced in \cite{YangLee1956}. Among the various hidden sector models, mirror matter has been extensively studied \cite{Kobzarev1966,Blinnikov1982,Berezhiani2004,Hodges1993,Foot2014,Hostert2023,Hao2022,Berezhiani2006,Berezhiani2019}. It is characterised by a particle content structurally identical to that of ordinary matter, with ordinary and mirror particles interacting only through gravitational and possibly weak interactions.
Within this framework, the ordinary and mirror sectors form parallel worlds, each governed by the same gauge symmetry group $G$. In the simplest scenario, $G$ corresponds to the gauge group of the Standard Model,
\[
G = G_{SM} \equiv SU(3)_c \otimes SU(2)_L \otimes U(1)_Y,
\]
which leads to an extended gauge structure of the form $G \otimes G$. A mirror Higgs doublet $\phi'$ is also introduced in addition to the standard Higgs doublet $\phi$. When the mirror symmetry is spontaneously broken \cite{Lee1974}, the vacuum expectation values of the Higgs fields are different, i.e. $\langle \phi \rangle \neq \langle \phi' \rangle$, leading to different mass spectra for the ordinary and mirror particles.
In this context, there exists a universal mixing mechanism between ordinary and mirror-neutral hadrons, which allows matter-mirror-matter oscillations due to spontaneous mirror symmetry breaking. For neutral hadrons such as neutrons, the physical states (neutron $n$ and mirror neutron $n'$) are not aligned with the mass eigenstates $n_1$ and $n_2$. As a result, a similar approach to that used for the ordinary two-flavour neutrino oscillation can be used to describe the $n-n'$ oscillation\cite{Berezhiani2006}.
The physical states $n$ and $n'$ are related to the mass eigenstates through a rotation: $\begin{pmatrix}
	n \\
	{n'}
\end{pmatrix}
=
\begin{pmatrix}
	\cos\theta & \sin \theta \\
	- \sin \theta & \cos \theta \\
\end{pmatrix}
\begin{pmatrix}
	n_1 \\
	n_2
\end{pmatrix}$, where $\theta $ is the mixing angle. The mass term of the Hamiltonian in the $n,n'$ basis, for a given spin polarization $s$, reads
\begin{equation}\label{H}
	H_s=\begin{pmatrix}
	m_n+\Delta E_s & \epsilon_{n n'} \\
	\epsilon_{n n'} & m_n+\delta m
	\end{pmatrix}
\end{equation}
where $m_n$ represents the neutron mass, $\epsilon_{n n'}$ is the mixing amplitude, and $\delta m = m_{n'} - m_n$ denotes the in-vacuum $n-n'$ mass splitting. The energy shift $\Delta E_s$ arises from neutron interactions with external fields, with the dominant contribution coming from the dipole coupling to an external magnetic field $\pmb{B}$. For spin polarized along $\hat{B}$, this shift is given by $\Delta E_s = -s\mu_{n}B$, where mixing is enhanced for spin aligned with the magnetic field ($s=1$). Focusing on this case, we define $\Delta E = -\mu_{n}B = |\mu_{n}B|$. It is important to note that the model parameters $\delta m$ and $\epsilon_{n n'}$ remain experimentally unmeasured, with current upper bounds of $\delta m \lesssim 10^{-7} \mathrm{eV}$ and $\epsilon_{nn'} \lesssim 10^{-9} \mathrm{eV}$ \cite{Berezhiani2019}.
The Hamiltonian \eqref{H} is diagonalized with a mixing angle and eigenvalues respectively 
\begin{equation}
	\tan\left( 2 \theta \right) = \frac{2 \epsilon_{n n'}}{\delta m - \Delta E}; \quad m_{1,2} = \frac{1}{2} \left( 2 m_n + \Delta E + \delta m \mp \sqrt{\left( \Delta E - \delta m \right)^2 + 4 \epsilon^2_{n n'}} \right).
\end{equation}
For a free particle with mass $m_j$ ($j=1,2$), the relativistic energy is $\omega_j = \sqrt{m_j^2 + k^2}$, where $k = |\pmb{k}|$ and $\pmb{k}$ is the 3-momentum. $\ket{n_j (t)} = e^{- i \omega_j t}\ket{n_j}$ gives the time evolution of the mass states, where $\ket{n_j} = \ket{n_j (t=0)}$. Omitting an irrelevant common phase factor, the physical states are the orthogonal combinations of
\begin{equation}\label{state2}
	\begin{aligned}
		\ket{n(t)}&=\cos{\theta} e^{ i \frac{\Delta \omega}{2} t} \ket{n_1} +\sin{\theta} e^{- i \frac{\Delta \omega}{2} t} \ket{n_2}\\
		\ket{n'(t)}&=-\sin{\theta} e^{ i \frac{\Delta \omega}{2} t} \ket{n_1} +\cos{\theta}  e^{- i \frac{\Delta \omega}{2} t} \ket{n_2}
	\end{aligned}
\end{equation}
where $\Delta \omega=\omega_2-\omega_1$.

\section{Geometric phase and experimental setup}
\begin{figure}[t]
	\centering
	\includegraphics[width=0.7\linewidth]{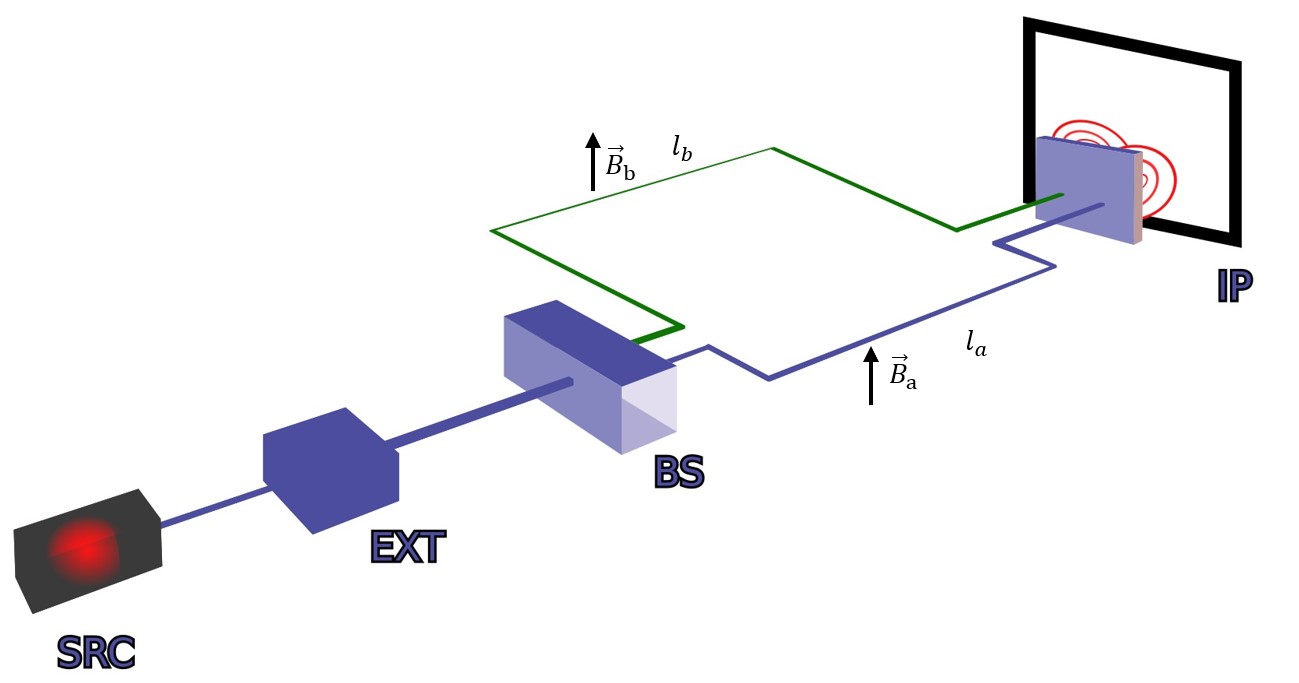}
	\captionsetup{font=small}
	\caption{(Color online). Schematic depiction of the interferometric apparatus. The beam is produced by a certain source \textbf{SRC}, polarized through an external device \textbf{EXT}. The beam then goes through a beam splitter \textbf{BS}. Interference is finally observed at the interference plane \textbf{IP}. }
	\label{fig:1}
\end{figure}

It is well known \cite{Geo4,Capolupo2018} that in the presence of particle mixing, an additional \emph{geometric phase} emerges in addition to the conventional dynamical phase associated with the time evolution. This phenomenon can be explicitly demonstrated by evaluating the phase acquired by the neutron in the context of $n-n'$ mixing. According to the kinematic definition introduced in \cite{Mukunda}, the geometric phase accumulated along a trajectory $\gamma$ is given by:
\begin{equation}
	\Phi^{g}(\gamma) = \Phi^{tot}(\gamma) - \Phi^{dyn}(\gamma),
\end{equation}  
where the total and the dynamical phase are respectively:  
\begin{equation}
	\Phi^{tot}(\gamma) = \arg \braket{\psi \left(\lambda_1\right)|\psi \left(\lambda_2\right)},  
\end{equation}  
\begin{equation}
	\Phi^{dyn} (\gamma) = \Im{\int_{\lambda_1}^{\lambda_2} { \braket{\psi \left(\lambda\right)|\dot{\psi} \left(\lambda\right)} d\lambda}},
\end{equation}  
where $\lambda \in [\lambda_1, \lambda_2]$ is an arbitrary parametrization of $\gamma$.  
For a neutron evolving under the action of the mixing Hamiltonian \eqref{H} from $t=0$ we have  
\begin{eqnarray}
	\nonumber && \Phi^{g} \left(t \right) = \arg \braket{n \left(0\right)|n \left(t\right)} - \Im{\int_{0}^{t} { \braket{n \left(t^\prime\right)|\dot{n} \left(t^\prime\right)} dt^\prime}} = \\
	&& \arg \left[ \cos\left({\frac{\Delta \omega}{2} t}\right)+i \sin\left({\frac{\Delta \omega}{2} t}\right) \cos{2 \theta } \right] - \frac{\Delta \omega t}{2} \cos{2 \theta }.
\end{eqnarray}  
The geometric phase $\Phi^{g}$ is both gauge-invariant and reparametrization-invariant, arising only from mixing effects and disappearing when the mixing angle $\theta$ is zero (i.e., when $\epsilon_{nn'} \to 0$). Thus, the measurement of $\Phi^{g}$ provides a direct test of the mirror-matter hypothesis and sets limits on $\epsilon_{nn'}$ and $\delta m$.
In neutron interferometry the measurable quantity is the total phase difference. By adjusting the arm lengths and the magnetic fields, the total phase difference can be made equal to the geometric phase difference. In this setup, the observed interference pattern comes entirely from the geometric phase and disappears in the absence of mixing.
We propose an experimental setup in which a neutron beam polarised along $\hat{y}$ is split into two sub-beams. These beams travel along interferometer arms of lengths $l_a$ and $l_b$ and experience magnetic fields $B_a$ and $B_b$, both aligned with $\hat{y}$. After recombination the interference pattern is observed. A schematic of the setup is shown in Fig.~\ref{fig:1}.
The difference in the dynamic phases must vanish at the interference plane to ensure that only the geometric phase contributes to the observed interference pattern. This condition is satisfied if the lengths of the two arms satisfy the relation: 
\begin{equation}\label{cond}
	l_b=\frac{\Delta \omega_a \cos{2\theta_a}}{\Delta \omega_b \cos{2\theta_b}} l_a,
\end{equation}  
where the mixing angles are $\theta_J = \frac{1}{2} \arctan \left(\frac{2\epsilon_{nn'}}{\delta m - |\mu_n B_J|} \right)$ and the energy differences $\Delta \omega_J = \omega_2 (B_J) - \omega_1(B_J)$,
explicitly depend on the magnetic field strengths for each arm, with $J = a, b$. Under these conditions, the difference in the geometric phases is left as the only contribution to the observed phase shift, which allows a direct probe of the phenomenon of neutron-mirror neutron mixing.
If the interferometric setup is arranged such that condition \eqref{cond} holds and if we express the time of flight as $t = \frac{l_a}{v}$, where $v$ is the neutron velocity, the observed phase difference reduces to:  
\begin{equation}\label{diff1}
	\begin{aligned}
		\Delta \Phi = \Delta \Phi^{g}  &=\arg \left[ \cos\left({\frac{\Delta \omega_a}{2 v} \frac{\cos{2\theta_a}}{ \cos{2\theta_b}} l_a}\right) + i \sin\left({\frac{\Delta \omega_a}{2 v} \frac{\cos{2\theta_a}}{ \cos{2\theta_b}} l_a}\right) \cos{2 \theta_2 } \right] -\\
		&-\arg \left[ \cos\left({\frac{\Delta \omega_a}{2 v}l_a}\right) + i \sin\left({\frac{\Delta \omega_a}{2 v}l_a}\right) \cos{2 \theta_a } \right].  
	\end{aligned}
\end{equation}

\section{Numerical analysis}
\begin{figure}
	\centering
	\includegraphics[width=0.7\linewidth]{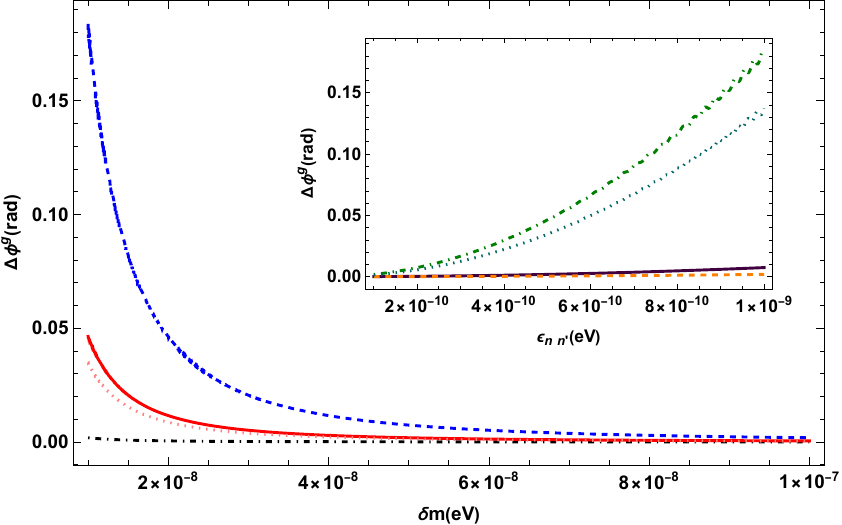}
	\captionsetup{font=small}
	\caption{(Color online).Geometric phase difference $\Delta \phi^g$ as a function of $\delta m$ for $\epsilon_{n n'} = 10^{-10} \ \mathrm{eV}$ (black dotted line), $\epsilon_{n n'} = 5 \times 10^{-9} \ \mathrm{eV}$ (red solid line), and $\epsilon_{n n'} = 10^{-9} \ \mathrm{eV}$ (blue dashed line). The inset shows $\Delta \phi^g$ as a function of $\epsilon_{n n'}$ for $\delta m = 10^{-8} \ \mathrm{eV}$ (green dotted line), $\delta m = 5 \times 10^{-8} \ \mathrm{eV}$ (purple solid line), and $\delta m = 10^{-7} \ \mathrm{eV}$ (orange dashed line). Parameters include neutron wavelength $\lambda = 10 \ \mathrm{\AA{}}$, magnetic fields $B_a = 0$, $B_b = 2 \times 10^{-4} \ \mathrm{T}$, and arm length $l_a = 20 \ \mathrm{cm}$. The dotted pink and cyan lines include the Earth's magnetic field $B_a = 5 \times 10^{-5} \ \mathrm{T}$, with $\epsilon_{n n'} = 5 \times 10^{-10} \ \mathrm{eV}$ (pink) and $\delta m = 10^{-8} \ \mathrm{eV}$ (cyan). }
	\label{fig:3}
\end{figure}

\begin{figure*}
	\centering
	\includegraphics[width=0.8\textwidth]{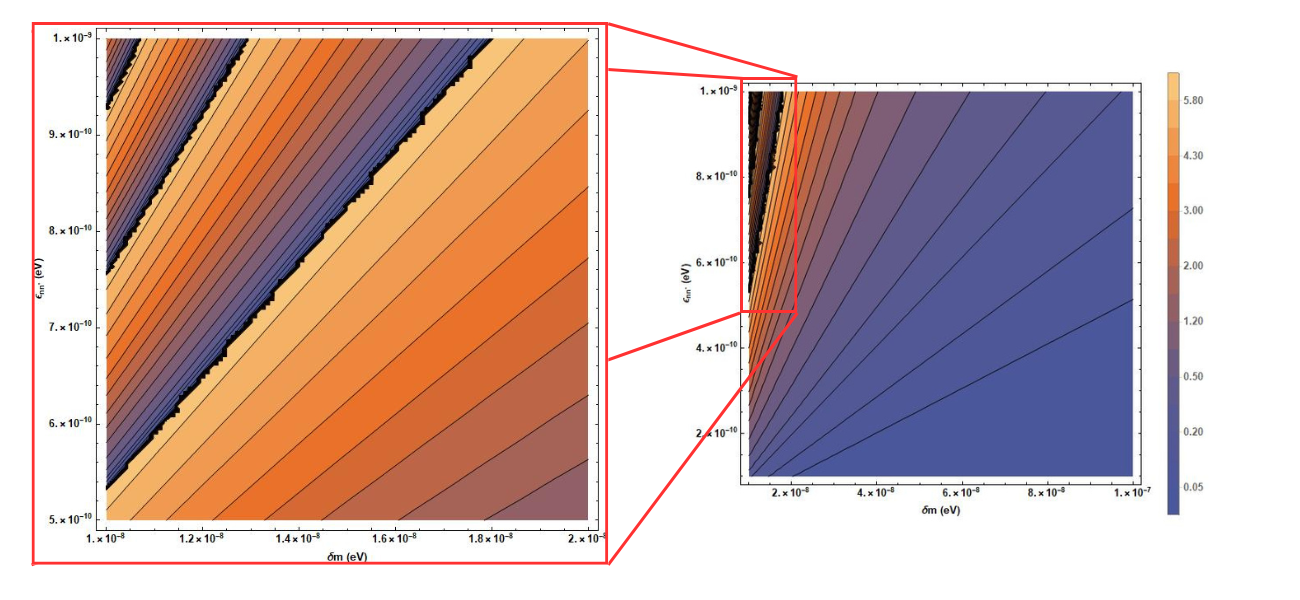}
	\captionsetup{font=small}
	\caption{(Color online). Geometric phase difference $\Delta \phi ^g$ modulo $2\pi$ (in radians) as a function of $\delta m$ and $\epsilon_{n n'}$ for values of the magnetic fields $B_a=2 \times 10^{-4} \ T$ and $B_a=2 \times 10^{-2} \ T$. Wavelength $\lambda = 10  \ \mathrm{\AA{}} $ and arm length $l_a = 20 \ \mathrm{cm}$ are considered.}
	\label{fig:5}
\end{figure*}

To illustrate the behaviour of the geometric phase difference in Eq. \eqref{diff1}, we adopt parameter values consistent with those used in modern neutron interferometry facilities. In these experiments, cold neutron beams - typically characterised by wavelengths $\lambda$ of the order of a few \AA{} - are used in interferometric setups with path lengths of several centimetres \cite{Facilities}. Fig. \ref{fig:3} shows a plot of $\Delta \Phi^{g}$, Eq. \eqref{diff1}, as a function of the parameters $\epsilon_{nn'}$ and $\delta m$, assuming an arm length of $l_a = 20\, \mathrm{cm}$, a neutron wavelength of $\lambda = 10\,\mathrm{\AA{}}$, and magnetic field values of $B_a = 0$ and $B_b = 2 \times 10^{-4}\,\mathrm{T}$.
\begin{figure}[htbp]
	\centering
	\captionsetup[subfigure]{labelformat=empty} 
	\begin{subfigure}{0.48\textwidth}
		\centering
		\includegraphics[width=\textwidth]{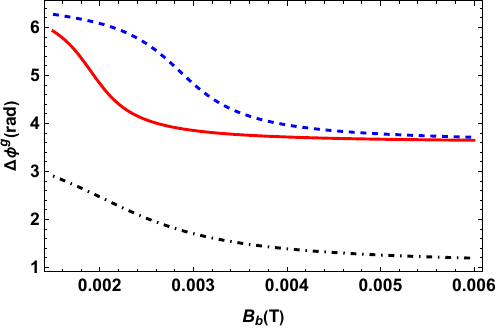}
		\caption{Figure 3\thesubfigure: (Color online). Geometric phase difference $\Delta \phi ^g$ modulo $2\pi$ as a function of $B_b$ for $\epsilon_{n n'}=10^{-10} \ \mathrm{eV}$ (the black dotted line), $\epsilon_{n n'}= 2 \times 10^{-10} \ \mathrm{eV}$ (the red line), $\epsilon_{n n'}= 3 \times 10^{-10} \ \mathrm{eV}$ (the blue dashed line) and $\delta m=0 $ ; we fix $l_a = 20 \ \mathrm{cm}$, $\lambda = 10 \ \mathrm{\AA{}}$ and $B_a = 5 \times 10^{-5} \ \mathrm{T}$..}
		\label{fig:6a}
	\end{subfigure} \hspace{0.01\textwidth}%
	\begin{subfigure}{0.48\textwidth}
		\centering
		\includegraphics[width=\textwidth]{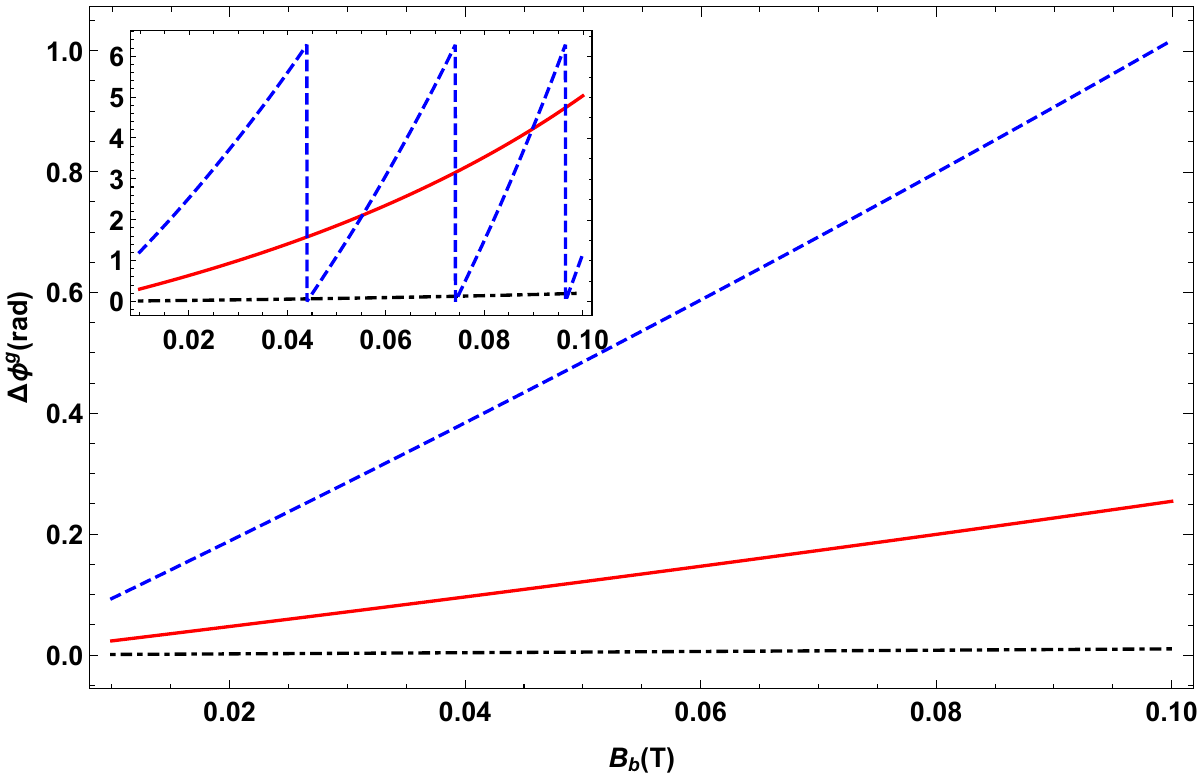}
		\caption{Figure 3\thesubfigure: (Color online). Geometric phase difference $\Delta \phi ^g$ modulo $2\pi$ as a function of $B_b$ for $\epsilon_{n n'}=10^{-10} \ \mathrm{eV}$ (the black dotted line), $\epsilon_{n n'}=5 \times 10^{-9} \ \mathrm{eV}$ (the red solid line) and $\epsilon_{n n'}=10^{-9} \ \mathrm{eV}$ (the dashed blue line); we fix $\delta m=10^{-7} \ \mathrm{eV}$( $\delta m=2 \times 10^{-8} \ \mathrm{eV}$ in the inset), $l_a = 20 \ \mathrm{cm}$, $\lambda = 10 \ \mathrm{\AA{}}$ and $B_a = 0$.}
		\label{fig:6b}
	\end{subfigure}
	
	\label{fig:6}
\end{figure}
The effect of the Earth's magnetic field is also analyzed in Fig. \ref{fig:3}. It can be seen that the geometric phase difference reaches higher values for smaller $\delta m$ and larger $\epsilon_{nn'}$. This behaviour can be explained by the mixing angle $\theta = \frac{1}{2} \arctan \left(\frac{2\epsilon_{n n'}}{\delta m - \Delta E}\right)$, which is maximised when $\delta m$ is small and $\epsilon_{nn'}$ is large. This trend is further highlighted in the contour plot of Fig. \ref{fig:5}, for certain magnetic field values.  
The geometric phase difference undergoes several full ($2\pi$) oscillations in the parameter space where $\epsilon_{nn'}$ is large and $\delta m$ is small. Since the mixing phenomenon is strongly enhanced in this region, it is of particular importance for experimental observations.\\
For the influence of magnetic fields a similar argument applies. The mixing effect is greatly enhanced when the energy shift, $|\mu_{n}B|$, approaches the resonance condition (i.e. when it is comparable to $\delta m$). In practice, this resonance is reached at about $B \sim 1\,\mathrm{T}$ for $\delta m \simeq 10^{-7}\,\mathrm{eV}$. As a result, the geometric phase difference can become large enough to complete multiple full oscillations as the magnetic field strength increases. This behaviour is illustrated in Fig. \ref{fig:6b}, where the inset shows that a smaller $\delta m$ leads to an increased mixing angle, producing multiple phase oscillations for $\epsilon_{nn'} = 10^{-9}\,\mathrm{eV}$.
Longer arms can increase the geometric phase difference, especially for small $\epsilon_{nn'}$ and large $\delta m$, where mixing is suppressed. Conversely, in regions of enhanced mixing, weaker magnetic fields and shorter arms are preferred to avoid excessive oscillations. Adjustable geometries and variable magnetic fields allow effective exploration of different parameter regimes.
Limitations to the observability of the phase difference of Eq. \eqref{diff1} are imposed by additional unwanted sources of phase shifts. The two most important sources in this case are curvature and spin precession. A certain curvature within at least one of the two arms is required to make the sub-beams meet at the interference plane, since $l_a \neq l_b$ is forced by the condition \eqref{cond}. While this inevitably adds an unwanted phase shift to at least one of the two paths, it is also the case that it can be easily controlled and properly subtracted to obtain the result of eq. \eqref{diff1}. A simple way to do this is to observe the interference pattern in the absence of magnetic fields, or for $B_a = B_b$, in which case \eqref{diff1} vanishes and the remaining phase shift is due solely to the geometry of the setup. Other minor sources of noise, such as the effect of mirrors and spurious interactions with matter, can be kept under control in the same way. An additional phase shift due to spin precession occurs if the neutron beam is not perfectly polarised with respect to the common direction of the magnetic fields $\hat{y}$ and a mixture of the two polarisations enters the magnetic field region. Since the intensities must be different $B_a \neq B_b$, an additional phase difference with respect to \eqref{diff1} would be observed. However, even this second limitation can be easily controlled by simply adjusting the arm lengths, depending on the spin composition of the beam, so that the magnetic phase shift is also removed.

\section{Conclusions}
The proposed experimental setup can explore, with minimal constraints, a substantial region of the parameter space associated with neutron-mirror neutron oscillations, detecting the geometric phase that inherently arises from this phenomenon.  
The proposed setup operates in a dichotomous manner: while it can confirm the presence of neutron-mirror neutron mixing, it cannot independently determine the individual parameters $\epsilon_{n n'}$ and $\delta m$ of the theoretical model. Nevertheless, when combined with other experimental techniques, the phase measurement can enable the determination of one parameter, provided the other is known.  
The observation of such a phase could unveil new physical scenarios, shedding light on a previously unobserved sector of particles and interactions. This sector might play a role in the dark sector of the universe and could be linked to experimental anomalies in particle physics.

\section*{Acknowledgements}
We acknowledge partial financial support from MUR and INFN, A.C. also acknowledges the COST Action CA1511 Cosmology and Astrophysics Network for Theoretical Advances and Training
Actions (CANTATA).

\end{document}